# Comprehensive evaluation of Mal-API-2019 dataset by machine learning in malware detection

Zhenglin Li[1, a], Haibei Zhu[2], Houze Liu[3], Jintong Song[4], Qishuo Cheng[5]

[1]Department of Computer Science and Engineering, Texas A&M University, USA;

[2]Computer Engineering, Georgia Institute of Technology, USA;

[3]Department of Computer Science and Engineering, New York University, USA;

[4]Department of Computer Science and Engineering, Boston University, USA;

[5]Department of Economics, University of Chicago, Chicago, USA.

[a]zhenglin_li@tamu.edu

**ABSTRACT**

This study conducts a thorough examination of malware detection using machine learning techniques, focusing on the evaluation of various classification models using the Mal-API-2019 dataset. The aim is to advance cybersecurity capabilities by identifying and mitigating threats more effectively. Both ensemble and non-ensemble machine learning methods, such as Random Forest, XGBoost, K Nearest Neighbor (KNN), and Neural Networks, are explored. Special emphasis is placed on the importance of data pre-processing techniques, particularly TF-IDF representation and Principal Component Analysis, in improving model performance. Results indicate that ensemble methods, particularly Random Forest and XGBoost, exhibit superior accuracy, precision, and recall compared to others, highlighting their effectiveness in malware detection. The paper also discusses limitations and potential future directions, emphasizing the need for continuous adaptation to address the evolving nature of malware. This research contributes to ongoing discussions in cybersecurity and provides practical insights for developing more robust malware detection systems in the digital era.

## 1. INTRODUCTION

As we navigate deeper into the digital age, cyberspace has continually transformed.[1] In the digital era, the escalating threat of malware poses a significant challenge to information security. Malware, designed to exploit and disrupt computer systems, has evolved into sophisticated forms, necessitating advanced detection methods. It also has a lot of applications in industry, including IoT security, edge device security, financial security and so on. [38-41] This paper undertakes a comprehensive analysis of malware detection techniques using the Mal-API-2019 dataset, aiming to compare their effectiveness and applicability, which will enable a diverse array of security analyses. [3-4]

Emphasizing the critical role of malware detection in safeguarding information systems, the paper outlines the evolution of malware and the need for robust detection strategies. It explores various methodologies, from traditional signature-based approaches to heuristic and behavior-based techniques, as well as the growing influence of machine learning and artificial intelligence in cybersecurity, especially, in recent years, the rise of machine learning and artificial intelligence providing stronger development momentum for related disciplines and industries. [44-47]

By examining these methodologies, the paper aims to elucidate their strengths, limitations, and adaptability to emerging threats. It provides insights into the nuances of malware detection, addressing aspects such as accuracy, speed, and resource requirements. Ultimately, this exploration aims to equip cybersecurity professionals, researchers, and policymakers with informed perspectives for enhancing malware detection systems in the face of evolving cyber threats.

## 2. RELATED WORK

### 2.1. Deep learning based Sequential model for malware analysis using Windows exe API Calls

In contrast to conventional statistical models such as native Bayesian, deep neural network (DNN) models have demonstrated remarkable efficacy in various scientific and engineering domains, notably in tasks like malware detection.[2][30] This paper introduces a deep learning-based sequential model tailored for malware analysis utilizing Windows exe API calls. The primary objective is to develop a classification method capable of categorizing malware types by analyzing their behavior. Sequential sentence classification is a natural language processing (NLP) technique that involves classifying a sequence of sentences into one or more categories or labels. [7-8] Notably, the research also contributes to the creation of a novel dataset specific to Windows operating systems, focusing on API calls. This dataset serves as a foundational resource for various studies in malware analysis and detection, including the one discussed in this paper. By leveraging deep learning techniques, the model offers a sophisticated approach to understanding and classifying malware behavior, thereby enhancing detection capabilities.

### 2.2. Understanding Random Forests: From Theory to Practice

This thesis provides a comprehensive examination of random forests, aiming to elucidate their underlying principles and practical implications. The research endeavors to dissect every aspect of the random forest algorithm, shedding light on its learning capabilities, internal mechanisms, and interpretability. A notable contribution of this work is its original complexity analysis of random forests, which demonstrates their favorable computational performance and scalability. Moreover, the thesis delves into implementation details, particularly within the context of Scikit-Learn, a widely used machine learning library. By offering a deep understanding of random forests, this research contributes valuable insights into their utility and effectiveness in various applications, including malware detection.

### 2.3. Advancements in Heuristic and Behavior-Based Techniques for Malware Detection

This research explores advancements in heuristic and behavior-based techniques for malware detection, aiming to enhance the efficacy of detection systems in identifying and mitigating cybersecurity threats. By analyzing the evolving landscape of malware, the study identifies the limitations of traditional signature-based approaches and highlights the need for more adaptive and proactive detection methods. Through a critical examination of heuristic and behavior-based techniques, the research aims to uncover innovative approaches to detecting malware based on its behavioral patterns and characteristics. By leveraging heuristic analysis, which focuses on identifying suspicious behavior rather than specific signatures, and behavior-based techniques that analyze the actions of programs in real-time, this research contributes to the development of more robust and agile malware detection systems.

## 3. DATA PRE-PROCESSING

Prior to engaging in trading activities, it is crucial to preprocess the data and eliminate any noise. [5-6] This research delves into the complex realm of malware analysis and detection, approaching it as a multi-classification problem. By recognizing the diverse nature of malware and its evolving tactics, the study aims to devise effective strategies for identifying and categorizing malicious software. To facilitate this endeavor, the research introduces a comprehensive dataset comprising Windows API calls associated with various types of malwares. This dataset serves as a foundational resource for understanding the behavioral patterns and characteristics of different malware strains.

The dataset has eight categories, including Trojan, Backdoor, Downloader, Worms, Spyware, Adware, Dropper, Virus. Among them, Backdoor / Trojan attacks have exposed the vulnerability of neural networks. [10-13]

In addition to presenting the dataset, our research outlines the methods employed for data pre-processing and feature representation. Recognizing the importance of data quality and representation in machine learning-based malware detection systems, the study emphasizes the need for robust pre-processing techniques. These techniques may include cleaning and formatting the data, handling missing values, and encoding categorical variables. Furthermore, the research explores feature representation methods tailored to the unique characteristics of malware datasets. This may involve extracting relevant features from the API call sequences, transforming them into numerical representations, and selecting informative features to feed into the classification models. [14]

By meticulously addressing the challenges of data pre-processing and feature representation, the research lays a solid foundation for subsequent analysis and model development. These preparatory steps are essential for ensuring the accuracy and effectiveness of machine learning algorithms in detecting and classifying malware. Moreover, by formulating malware detection as a multi-classification issue and leveraging advanced data processing techniques, the study aims to contribute to the ongoing efforts in cybersecurity to combat the ever-evolving threat landscape posed by malicious software.

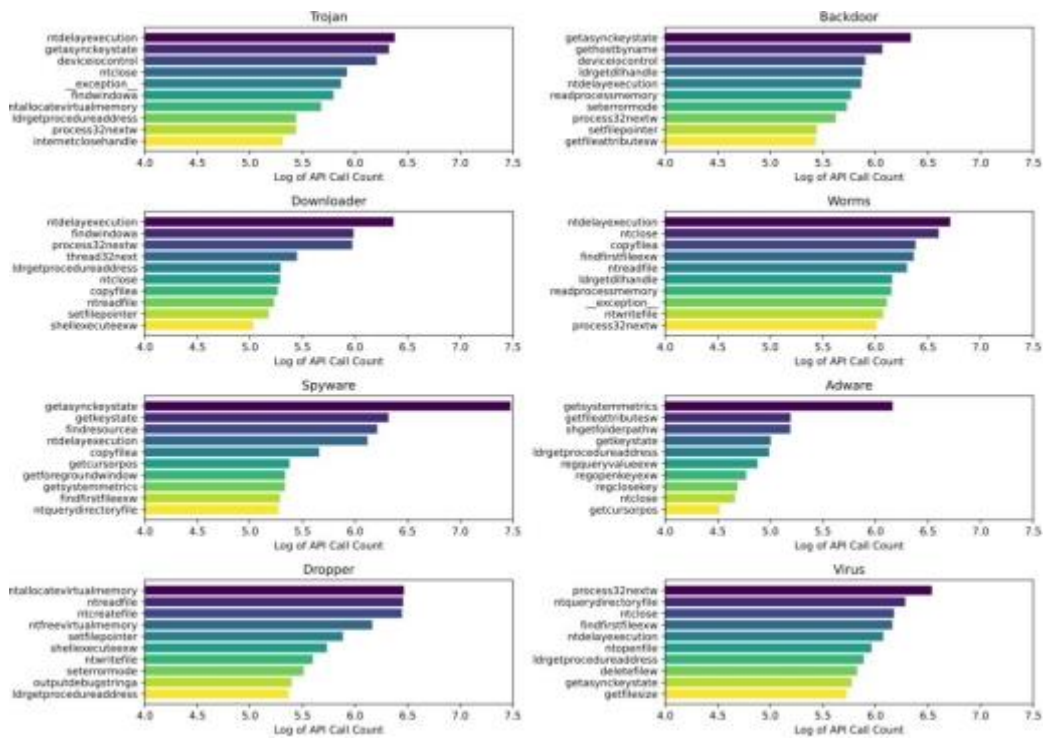

**Figure 1:** Top 10 Frequent Windows API Calls in Each Malware Types

## 3.1. Windows API Calls

The Windows API serves as a fundamental interface for programmers to interact with the operating system at a low level. It offers functions categorized into eight distinct groups, including Base Services, Advanced Services, Graphics Device Interface, and more. Malware often leverages these API calls to gain access to and manipulate restricted resources, such as file operations, registry manipulations, and network communications. Therefore, a thorough analysis of Windows API calls is crucial for understanding and detecting malware.

## 3.2. Dataset

The Mal-API-2019 dataset, created using Cuckoo Sandbox, is a valuable resource for cybersecurity research, particularly in malware analysis. This dataset focuses on analyzing Windows OS API calls and is structured as a CSV file, making it suitable for machine learning applications. It encompasses eight primary malware categories and provides insights into the API calls associated with each category. Visualizations, such as Fig. 1, highlight the distribution of API calls across different malware types.

## 3.3. Embedding Techniques

Frequency embedding and temporal embedding are employed to capture contextual and time-dependent patterns in the dataset. While frequency embedding focuses on the usage frequency of API calls, temporal embedding captures temporal patterns in the data. However, experiments show that these embedding methods yield similar results compared to TF-IDF embedding, indicating limited improvement in model performance.

## 3.4. TF-IDF Embedding

TF-IDF (Term Frequency-Inverse Document Frequency) embedding is utilized to generate numerical representations of API call sequences. This strategy emphasizes rare API calls that are indicative of unique or suspicious behavior, while reducing the impact of common, less informative calls. TF-IDF embedding enhances the model's ability to identify malware by highlighting significant patterns in the data.

## 3.5. Principle Component Analysis (PCA)

The dataset contains a large number of unique API calls, resulting in a high-dimensional feature space. To manage this complexity, Principal Component Analysis (PCA) is employed to transform the original features into a new set of uncorrelated variables known as principal components. By focusing on components that explain the most variance, PCA helps filter out noise and irrelevant information, aiding in the detection of malicious patterns in malware.

# 4. MALWARE CLASSIFICATION MODELS

In this section, we discuss the models used for multi-class classification. categorizing them into two types: ensemble and non-ensemble models. For the ensemble models, we utilized Random Forest and XGBoost. Regarding the non-ensemble models, we implemented K Nearest Neighbor (KNN) and Neural Networks. This classification provides a comprehensive understanding of different approaches in machine learning for effectively handling multi-class classification tasks.

## 4.1. Random Forest

The Random Forest model, a classic application of bagging consisting of multiple independent decision trees, has been widely recognized for its effectiveness in various machine learning tasks, including malware detection. As detailed by Louppe (2015) in Understanding Random Forests: From Theory to Practice, Random Forest leverages the strength of multiple decision trees to produce a more robust and accurate model.[9] This is particularly beneficial in malware detection where the complexity and variability of data are high. Louppe's work provides indepth insights into the inner workings of Random Forest, shedding light on its learning capabilities and interpretability, which are crucial for understanding the model's decisions in the context of cybersecurity threats. [15]

As a classic application of bagging, Random Forest has the following two important features:

Bootstrap: Given a dataset, instead of training one decision tree classifier on the entire dataset, we sample with replace- ment to create different datasets, called bootstraps, and train a classification or regression model on each of these bootstraps. This step ensures each bootstrap is independent of its peers.

Aggregating: The most common aggregation method is majority voting for classification tasks (selecting the class with the most votes). [16]

The base model of random forest is decision tree, thus, when build the model, we need to consider hyperparemeters like splitting criteria, minimum samples in leaf, etc. into account.

## 4.2. XGBoost

XGBoost, or Extreme Gradient Boosting, is an advanced implementation of gradient boosting algorithms. Unlike bag- ging, which trains the base models independently, boosting focuses on iteratively improving the base models by giving more weight to data points that are difficult to classify cor- rectly. Each time base models trained on the same dataset but with different weight. Its core principle is to build new models that predict the residuals or errors of prior models and then combine them to make the final prediction. The objective function is shown as follows.

$$Obj(\theta) = \sum l(y, \hat{y}) + \sum \Omega(f) \quad (1)$$

In equation (1), Obj($\theta$) is the objective function to be min-imized. $l(y_i, \hat{y}_i)$ represents the loss function, which measures the difference between the predicted value $\hat{y}_i$ and the actual value $y_i$. $\Omega(f_k)$ is the regularization term, which helps to smooth the final learned weights to avoid overfitting. K is the number of trees. $\theta$ represents the parameters of the model. By iteratively optimizing the objective function, the model's performance is improved.

## 4.3. K Nearest Neighbour

K-Nearest Neighbors (kNN) is a simple yet effective algorithm utilized in machine learning for classification tasks. It operates on the principle that similar data points tend to cluster together in a feature space. The algorithm finds the k nearest neighbors to a given query point based on a specified distance metric, such as Euclidean distance, and makes predictions based on the majority class among these neighbors. In the context of malware detection, kNN can be employed to classify samples based on similarities in their feature representations, such as API call sequences.

### 4.4. Neural Networks

As the importance of artificial intelligence (AI) and machine learning (ML) continues to grow in the realm of cybersecurity, neural networks have emerged as a powerful tool for malware detection. In this study, we implement a 4-layer neural network architecture for classification tasks. Each layer consists of a fully connected neural network followed by Rectified Linear Unit (ReLU) activation and dropout layers to prevent overfitting. The labels are encoded using One Hot Encoding, and the output layer comprises 8 neurons to correspond to the eight classes for classification. The Adam optimizer and categorical cross-entropy loss function are chosen for model training. This approach leverages the expressive power of neural networks to learn intricate patterns in the dataset, contributing to more accurate and robust malware detection systems. But, it also requires a lot of training data and may pose a threat to the security and privacy of the training data and with an issue of insufficient explainability. [17-19]

## 5. EXPERIMENTS

### 5.1. Setups

1) Environment: The experiments were conducted using Kaggle Cloud Notebook or Alibaba Cloud Elastic Compute Service. [20] The setup included the following specifications:

- Platform processor: x86-64 machine
- System info: Linux
- Python compiler: GCC 12.3.0
- Python version: 3.10.12
- Libraries: Scikit-learn 1.2.2, Pandas 2.0.3, Tensorflow 2.13.0.

2) Model Setup: Each model's performance was evaluated using 5-fold cross-validation, and the model's hyperparameters were optimized using grid search.

### 5.2. Results

The evaluation of different machine learning models for malware detection, summarized in Table I, provided valuable insights. These results were obtained through rigorous 5-fold cross-validation and hyperparameter optimization.

### TABLE I: Model Performance

| Model | Avg. ACC | Precision | Recall | F1 |
|---|---|---|---|---|
| KNN | 0.54 | 0.55 | 0.56 | 0.55 |
| Random Forest | 0.68 | 0.68 | 0.69 | 0.69 |
| XGBoost | 0.68 | 0.70 | 0.69 | 0.69 |
| Neural Networks | 0.56 | 0.58 | 0.51 | 0.52 |
| KNN + PCA | 0.54 | 0.54 | 0.56 | 0.55 |
| RF + PCA | 0.62 | 0.64 | 0.64 | 0.64 |
| XGB + PCA | 0.62 | 0.64 | 0.63 | 0.63 |
| KNN[a] | 0.36 | 0.35 | 0.35 | 0.34 |
| Random Forest[a] | 0.47 | 0.46 | 0.47 | 0.46 |

### 5.2.1. Performance Comparison:

The table shows the average accuracy, precision, recall, and F1 score of various models on the Mal-API-2019 dataset. Among the models, Random Forest and XGBoost demonstrated superior performance, achieving an average accuracy of 0.68. This aligns with existing literature, highlighting the effectiveness of ensemble methods in handling complex, high-dimensional data like malware signatures.

K Nearest Neighbor (KNN) exhibited a relatively lower average accuracy of 0.54, possibly due to its sensitivity to high dimensionality and noise in the dataset. The Neural Networks model, although pending completion of its evaluation metrics, showed an average accuracy of 0.56, indicating potential but also the need for further optimization.

### 5.2.2. Precision and Recall Analysis:

XGBoost slightly outperformed Random Forest in terms of precision, with a value of 0.70 compared to Random Forest's 0.68. This suggests that XGBoost was slightly better at minimizing false positives. Higher recall stands for less malware that the detection system will miss. [21-22] XGBoost and Random Forest models demonstrated similar recall rates, indicating their effectiveness in identifying true malware instances.

### 5.2.3. Implications of Findings:

The comparable performance of Random Forest and XGBoost underscores the effectiveness of ensemble methods in malware detection. The slightly higher precision of XGBoost may be crucial in contexts where false positives are particularly undesirable. The lower performance of KNN and Neural Networks highlights the challenges these models face in dealing with the complex and varied nature of malware signatures.

### 5.2.4. Surprising Trends:

The close performance of Random Forest and XGBoost was somewhat unexpected, given XGBoost's reputation for outperforming other models in similar tasks. This suggests that factors like training time and model interpretability may influence the choice between these models in the specific context of malware detection.

### 5.2.5. Comparison with Existing Methods:

Our findings are consistent with current research, emphasizing the efficacy of ensemble methods in malware detection. However, our study provides new insights into the relative performance of these methods compared to KNN and Neural Networks, specifically on the Mal-API-2019 dataset.

## 6. CONCLUSION

This paper provides a comprehensive exploration of machine learning techniques for malware detection, with a focus on leveraging the Mal-API-2019 dataset. Our investigation encompasses a range of methodologies, including ensemble models like Random Forest and XGBoost, as well as non-ensemble models such as K Nearest Neighbor and Neural Networks. Through a comparative analysis, we uncover valuable insights into the effectiveness of these models in detecting malware.

The results highlight the superior performance of ensemble models, particularly Random Forest and XGBoost, in terms of accuracy, precision, and recall. Their robustness in handling complex data structures and ability to capture intricate patterns in malware behavior contribute to their effectiveness. However, non-ensemble models still offer valuable insights, especially in scenarios prioritizing model interpretability and computational efficiency.

Moreover, our study underscores the crucial role of data preprocessing techniques, such as TF-IDF representation and Principal Component Analysis, in enhancing model performance. These techniques aid in distilling the dataset to its most informative features, facilitating more accurate malware classification.

Despite these advancements, our research has limitations. The scope of the dataset is confined to Windows API calls, potentially missing other malware attributes. And the method presented in this paper cannot predict unseen malware, which is also a common drawback on machine learning approach.[23] Additionally, the dynamic nature of malware evolution requires continuous updates to datasets and models to maintain effectiveness.

Future research directions could focus on expanding the dataset to include diverse malware signatures and integrating deep learning techniques, for example, long short-term memory (LSTM) models, which have shown promise in other NLP tasks. [24-29] Exploring real-time detection systems, more advanced deep learning models, and investigating model robustness against adversarial attacks are promising areas for further exploration. [30-31] Additionally, learning from demonstrations and reinforcement learning has shown good results. [42-43] What's more, large language models (LLMs) and multimodal systems can also help. [32-37]

In conclusion, this study significantly contributes to the field of cybersecurity by providing a nuanced understanding of machine learning models in malware detection. By continuing to advance these techniques and addressing future research avenues, we can enhance our capabilities in combating evolving cyber threats effectively.

## ACKNOWLEDGEMENT

They say that good premises do not guarantee good stories, and it holds true for this journey as well. The successful conclusion of this paper owes a debt of gratitude to the invaluable assistance, unwavering support, and trust extended by colleagues, friends, and family. The authors express heartfelt appreciation to the anonymous reviewers whose constructive feedback played a crucial role in shaping the outcome.